\documentstyle[epsf]{europhys}
\def\And{{\rm and\ }}

\def\stars{\bigskip\centerline{***}\medskip}

\newif\ifboo \boofalse

\begin{document}

\euro{0}{0}{0-0}{1998}
\Date{26 March 1998}
\shorttitle{O. THEISSEN et. al. LATTICE-BOLTZMANN MODEL OF AMPHIPHILIC SYSTEMS}

\title{Lattice-Boltzmann Model of Amphiphilic Systems}          

\author{O.Theissen\inst{1}, G. Gompper\inst{1} \And D. M. Kroll\inst{2}}

\institute{
  \inst{1} Max-Planck-Institut f\"ur Kolloid- und Grenzfl\"achenforschung -\\ 
  Kantstr. 55, 14513 Teltow, Germany\\
  \inst{2} Department of Medicinal Chemistry \And
  Minnesota Supercomputer Institute, \\
  University of Minnesota -  
  308 Harvard Street SE, Minneapolis, MN 55455, USA}

\rec {}{}

\pacs{
  \Pacs{47}{20.Ma}{Interfacial instability}
  \Pacs{82}{70.$-$y}{Disperse systems}
  \Pacs{83}{10.Lk}{Multiphase flows}
}

\maketitle

\begin{abstract}
A lattice-Boltzmann model for the study of the dynamics of
oil-water-surfactant mixtures is constructed.  The model, which is
based on a Ginzburg-Landau theory of amphiphilic systems with a
single, scalar order parameter, is then used to calculate the spectrum
of undulation modes of an oil-water interface and the spontaneous
emulsification of oil and water after a quench from two-phase
coexistence into the lamellar phase.  A comparison with some
analytical results shows that the model provides an accurate
description of the static and dynamic behavior of amphiphilic systems.
\end{abstract}

\section{Introduction}
Self-assembling amphiphilic system are characterized by a large
variety of phases \cite{gg:gomp94d,gelb94}.  Many of these are
structured on mesoscopic length scales.  Examples include the swollen
lamellar phase, the microemulsion phase in ternary amphiphilic
mixtures, and the sponge phase in aqueous amphiphile solutions.  In
all of these cases, the amphiphile assembles into two-dimensional
sheets which separate space into two networks of oil- and/or
water-channels, both of which span the whole system.

The mesoscopic structure leads to dynamical behavior which differs
qualitatively from that of simple fluid mixtures.  In addition to the
diffusive dynamics of all components, hydrodynamic flow plays an
important role.  Three classes of theoretical models which include
hydrodynamic effects have been employed so far.  The first are
time-dependent Ginzburg-Landau
models \cite{gg:gomp94a,gg:gomp96d,paet95_96b} which are coupled to a
linearized Navier-Stokes equation ({\it i.e.} generalizations of model
H of Hohenberg and Halperin \cite{hohe77} to amphiphilic systems).
The second are membrane models, in which the hydrodynamics of the
solvent is taken into account \cite{broc75,mess90}.  Finally, a
lattice-gas model \cite{bogh96,emer97a} --- similar in spirit to
vector lattice models employed to study equilibrium properties of
microemulsions \cite{gg:gomp94d} --- has been formulated and applied
to study the dynamics of these systems.

An interesting alternative to these descriptions of the hydrodynamics
of complex fluids is the Lattice-Boltzmann (LB)
approach \cite{mcna88,qian92,guns91,shan93}.  It has recently been
shown how LB methods can be generalized to study the dynamics of
non-ideal fluids \cite{swif95,swif96} and fluid
mixtures \cite{orla95,swif96,gonn97} with a given free-energy
functional.  In this paper we show how this approach can be extended
to study the dynamical behavior of amphiphilic systems.

\section{Lattice-Boltzmann Approach}
The basic variables of a LB model for two-component fluid mixtures are
the ``microscopic'' distribution functions $f_i({\bf r})$ and
$g_i({\bf r})$, which are related to the density $\rho$, density
difference $\Delta \rho$, and velocity ${\bf u}$ by
\begin{equation} \label{observables}
\rho({\bf r}) = \sum_i f_i({\bf r}); \ \ \ 
\Delta\rho({\bf r}) = \sum_i g_i({\bf r}); \ \ \ 
\rho({\bf r}) {\bf u}({\bf r}) = \sum_i f_i({\bf r}) {\bf e}_i \ ,
\end{equation}
where $i$ denotes the possible velocity states, ${\bf e}_0=0$, and 
${\bf e}_i$ with $i\ge 1$ are the unit lattice vectors.
The dynamics of the system is determined by the temporal evolution 
equations

\begin{eqnarray} \label{f_LB}
f_{i}({\bf r}+{\bf e}_{i},t+1) &=& f_{i}({\bf r},t) - 
       \frac{1}{\tau_\rho}[f_{i}({\bf r},t)-f_{i}^{eq}({\bf r},t)] \\
\label{g_LB}
g_{i}({\bf r}+{\bf e}_{i},t+1) &=& g_{i}({\bf r},t) - 
       \frac{1}{\tau_\Delta}[g_{i}({\bf r},t)-g_{i}^{eq}({\bf r},t)] \ .
\end{eqnarray}
It has been shown in ref.~\cite{orla95} that the time dependence of
$\Delta \rho({\bf r})$, $\rho({\bf r})$, and the momentum density is 
described by Cahn-Hilliard and
Navier-Stokes equations on sufficiently large length scales when the
zeroth and first moments of $f_i^{eq}$ and $g_i^{eq}$ are required to
fulfill eq.~(\ref{observables}), and, in addition
\begin{eqnarray} \label{nonideal_f}
\sum_i f_i^{eq} e_{i\alpha} e_{i\beta} &=& P_{\alpha\beta} 
                          + \rho u_\alpha u_\beta  \\
\label{nonideal_g}
\sum_i g_i^{eq} e_{i\alpha} e_{i\beta} &=& 
   \Gamma \,
    \Delta\mu
           \, \delta_{\alpha\beta} + \Delta\rho \, u_\alpha u_\beta \ ,
\end{eqnarray}
where $\Delta\mu=\delta{\cal F}[\rho,\Delta\rho] / \delta
  \Delta\rho$ is the chemical potential for the density difference
$\Delta\rho$, $\Gamma$ is the mobility, and $P_{\alpha\beta}$ the
pressure tensor.  We choose $\tau_\Delta = (1+1/\sqrt{3})/2$ to
minimize terms which break Galilean invariance \cite{orla95}.

\section{Amphiphilic Systems}
Many equilibrium properties of ternary amphiphilic systems can be
accurately described using a Ginzburg-Landau model for a single, scalar
order parameter $\phi({\bf r})$, which we identified with the local
{\it density} difference, $\Delta \rho({\bf r})$, of oil and water.
This model is defined by the free energy functional
\cite{gg:gomp90d,gg:gomp94d}
\begin{equation} \label{GL_phi}
{\cal F}_{\phi}[\phi]
= \int \,d^dr \bigl\{ c(\nabla^2\phi)^{2} 
+ g(\phi)(\nabla\phi)^{2} + f(\phi) \bigr\}\ . 
\end{equation}
We assume that the full free-energy functional, which includes the 
density field, can be written as 
${\cal F}[\rho,\phi] = {\cal F}_{\phi}[\phi] + {\cal F}_{\rho}[\rho]$,
where ${\cal F}_{\rho}[\rho] = \int \,d^dr \bigl\{ b(\nabla\rho)^{2} + 
\psi(\rho) \bigr\}$.
A similar model, with $g(\phi)=g_0=\;$const.  in eq.~(\ref{GL_phi}) and
a different ansatz for the free energy density $f(\phi)$ (see below),
has been employed in ref.~\cite{gonn97} to study the spinodal
decomposition to the lamellar phase.

The first task in the construction of the LB model is to determine the
pressure tensor $P_{\alpha\beta}$, which is defined by the condition
of hydrostatic equilibrium
\begin{equation} \label{pressure}
\partial_\beta P_{\alpha\beta}({\bf r}) = 
      \rho({\bf r})\partial_\alpha\mu({\bf r}) 
          +\Delta\rho({\bf r})\partial_\alpha \Delta\mu({\bf r})\ , 
\end{equation}
in a system with external potential $\mu({\bf r})$ for the density
$\rho({\bf r})$. The pressure tensor is not unique, since
eq.~(\ref{pressure}) is invariant under the substitution
$P_{\alpha\beta}\to P_{\alpha\beta}+G_{\alpha\beta}$ with
$\partial_{\beta}G_{\alpha\beta}=0$. Indeed, a Chapman-Enskog expansion 
shows that only the divergence
of the pressure tensor appears in the Cahn-Hilliard and Navier-Stokes
equations.  Eq.~(\ref{pressure}) can be solved by expanding in powers
of $\phi({\bf r}) \equiv\Delta\rho({\bf r})$ and its derivatives.
After some algebra, we obtain
\begin{equation} \label{tensor}
P_{\alpha\beta} = p_0 \delta_{\alpha\beta} + {\tilde P}_{\alpha\beta}\ , 
\end{equation}
where 
\begin{eqnarray} \label{p_0}
p_{0} &=& 2c\phi(\nabla^2\nabla^2\phi) - c(\nabla^2\phi)^{2} 
     - 2g(\phi)\phi\nabla^2\phi - g(\phi)(\nabla\phi)^{2} 
     - g'(\phi)\phi(\nabla\phi)^{2} \nonumber \\
  && - 2b\rho\nabla^2\rho - b(\nabla\rho)^{2} + \phi f'(\phi) - f(\phi) 
     + \rho\psi'(\rho)- \psi(\rho) 
\end{eqnarray}
is the (isotropic) local thermodynamic pressure, and
\begin{eqnarray} \label{P_tilde}
\tilde P_{\alpha\beta} &=& 2b(\partial_{\alpha}\rho)(\partial_{\beta}\rho)
     + 2g(\phi)(\partial_{\alpha}\phi) (\partial_{\beta}\phi) 
     + c \Lambda (\nabla^2\phi)^{2}\delta_{\alpha\beta} 
     - 2 c (\partial_{\alpha}\partial_{\gamma}\phi)
                      (\partial_{\beta}\partial_{\gamma}\phi) \nonumber \\
  && + c(\Lambda-2)\bigl\{(\partial_{\alpha}\phi)
            (\partial_{\beta}\nabla^2\phi) + 
     (\partial_{\beta}\phi)(\partial_{\alpha}\nabla^2\phi)\bigr\} 
     + 2c(1-\Lambda)(\partial_{\alpha}\partial_{\beta}
                 \partial_{\gamma}\phi)(\partial_{\gamma}\phi) \nonumber \\
  && + c\Lambda\phi(\partial_{\alpha}\partial_{\beta}\nabla^2\phi) 
     - c\Lambda\phi(\nabla^2\nabla^2\phi)\delta_{\alpha\beta}  
     + c(4-\Lambda)(\partial_{\alpha}\partial_{\beta}\phi)
                                                (\nabla^2\phi)  \ ,
\end{eqnarray}
with an arbitrary real number $\Lambda$, which reflects the
non-uniqueness of the pressure tensor. For $c=0$ and
$g(\phi)=g_0$ eq.~(\ref{P_tilde}) reduces to the well known expression
for inhomogeneous liquids~\cite{evan79}.

Since this result is rather lengthy, it is important to see that
eqs.~(\ref{observables}) and (\ref{nonideal_f}) can be solved
explicitly for {\it any} pressure tensor.  In the case of the
two-dimensional triangular lattice, the equilibrium distribution
$f^{eq}$ is found to be
\begin{eqnarray} \label{f_eq0}
  f_{0}^{eq} &=& \rho - Tr P_{\alpha\beta} - \rho {\bf u}^{2} \\
\label{f_eqi}
  f_{i}^{eq} &=& -\frac{1}{6} Tr P_{\alpha\beta} 
              + \frac{\rho}{3}({\bf e}_{i}\cdot{\bf u}) 
              - \frac{\rho}{6}{\bf u}^{2} 
              + \frac{2}{3}\rho ({\bf e}_{i}\cdot {\bf u})^{2} 
              + \frac{2}{3}P_{\alpha\beta} e_{i\alpha} e_{i\beta} \ .
\end{eqnarray}
The distribution $g^{eq}$ can be obtained from
eqs.~(\ref{f_eq0}) and~(\ref{f_eqi}) for $f^{eq}$ by replacing
$P_{\alpha\beta}$ by $\Gamma(\delta \,{\cal F}[\rho,\Delta\rho]/\delta
\,\Delta\rho) \, \delta_{\alpha\beta}$.  It can be shown easily that this
result agrees with the solutions of the special cases studied in
refs.~\cite{swif95,orla95,swif96}.

\section{Equilibrium Properties}
To test the correctness of the approach described above, we have
followed refs.~\cite{swif95,orla95,swif96} and first studied the
equilibrium properties of our model.  We have checked both the density
profiles and the phase diagram for the Ginzburg-Landau model
introduced in ref.~\cite{gg:gomp92i}, with $f(\phi)= \omega_0
(\phi^2-\phi_b^2)^2 (\phi^2+f_0)$ and $g(\phi) = g_0 + g_2 \phi^2$,
and for a simple ideal-gas free-energy density $\psi(\rho) =
T\rho\ln(\rho)$.  The deviation of the LB results from
the solutions of the mean-field equations is on the order of a few
percent if the parameters are chosen such that the interface width is
at least several lattice constants.  The phase diagram is determined
by studying the stability of the microemulsion phase with respect to
small density fluctuations.  Similarly, the phase transition between
the lamellar phase and oil-water coexistence is located by looking for
either the growth of lamellae or a coarsening of the bicontinuous
oil-water network after a quench from the microemulsion phase.  In
both cases, the transitions are found to agree within a few percent
with those of the mean-field phase diagram.

\begin{figure}
\epsfxsize=14cm
\epsffile{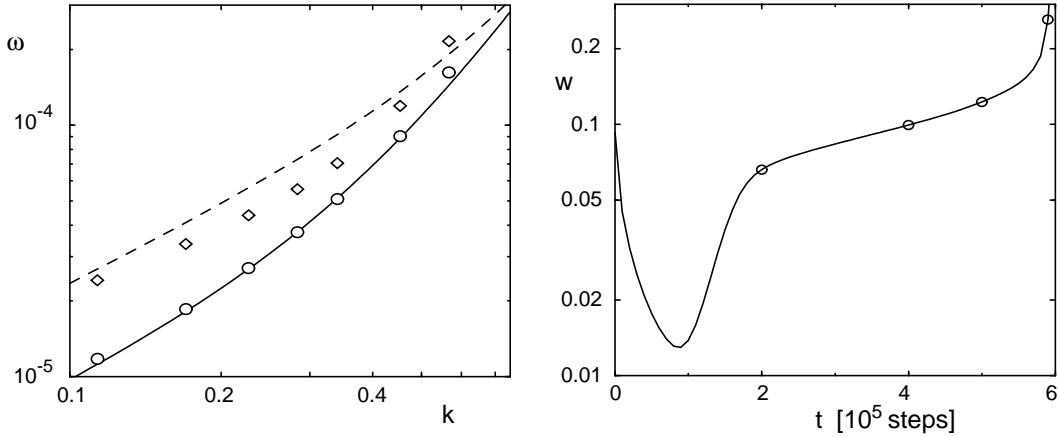}
\caption{
Spectrum of undulation modes.  The parameters are
$c=0.01$, $\omega_b=0.01$, $\phi_b=0.5$, $f_0=0.1$,
$g_2=4\phi_b(\sqrt{c\omega(1+4f_0)} - g_0)+0.01$, $\Lambda=0.123$,
$b=0.05$, $T=0.5$, $\Gamma=0.1$, and $\tau_\rho=5.0$.  Results are
shown for $g_0=+0.004$ ($\diamond$, dashed line) and $g_0=-0.004$
($\circ$, full line), which correspond to large and small $\sigma$,
respectively.}
\label{fig1}
\caption{
Interface width $w$ after a quench of a system from oil-water
coexistence into the lamellar phase.  The parameters are the same as
in fig.~\ref{fig1}, except for $\tau_\rho=1.0$ and $g_0=-0.0125$.  The
symbols indicate times for which the interfacial tension profile is
shown in fig.~\ref{fig4}.}
\label{fig2}
\end{figure}

\section{Undulation Modes}
Due to the presence of the amphiphile, interfaces in amphiphilic
systems can have an ultralow interfacial tension $\sigma$.  The
fluctuations of an interface are controlled in this limit by the
bending rigidity $\kappa$.  For the typical length scales of
amphiphilic systems, the hydrodynamic flow is usually in the
low-Reynolds-number, creeping-flow regime.  In this case, a
calculation of membrane fluctuations in the Stokes approximation
\cite{broc75} shows that the spectrum of the undulation modes of
wavenumber $k$ is given by
\begin{equation} \label{spectrum}
\omega(k) = \frac{i}{4\eta}\Bigl(\sigma+\kappa k^{2}\Bigr) k \ ,
\end{equation}
where $\eta$ is the shear viscosity.  We have determined the
relaxation times of the undulation modes of different wavelengths from
our LB model of lattice sizes $64\times 128$ and $128 \times 128$ with
periodic boundary conditions.  The results are shown in
fig.~\ref{fig1} for two values of the parameter $g_0$, which we use to
tune the interfacial tension \cite{gg:gomp92i}.  One of the big
advantages of the LB approach used here is that {\it all} the
coefficients in eq.~(\ref{spectrum}) can be determined independently
--- $\sigma$ and $\kappa$ from a study of the elasticity of curved
interfaces in the Ginzburg-Landau model \cite{gg:gomp92i,gg:gomp94d},
and $\eta=(2\tau_\rho-1)/8$ from a Chapman-Enskog expansion of the
Boltzmann equation (\ref{f_LB}) \cite{swif96}.  The simulation results
can therefore be compared directly with theory.  It can be seen from
fig.~\ref{fig1} for two values of the interfacial tension that the
agreement is very good; the almost perfect agreement for $g_0=-0.004$
is fortuitous, however.

We have checked explicitly that our results do {\it not} depend on
the parameter $\Lambda$ in the pressure tensor, eq.~(\ref{P_tilde}),
as they should.

\begin{figure}
\epsfxsize=14cm
\epsffile{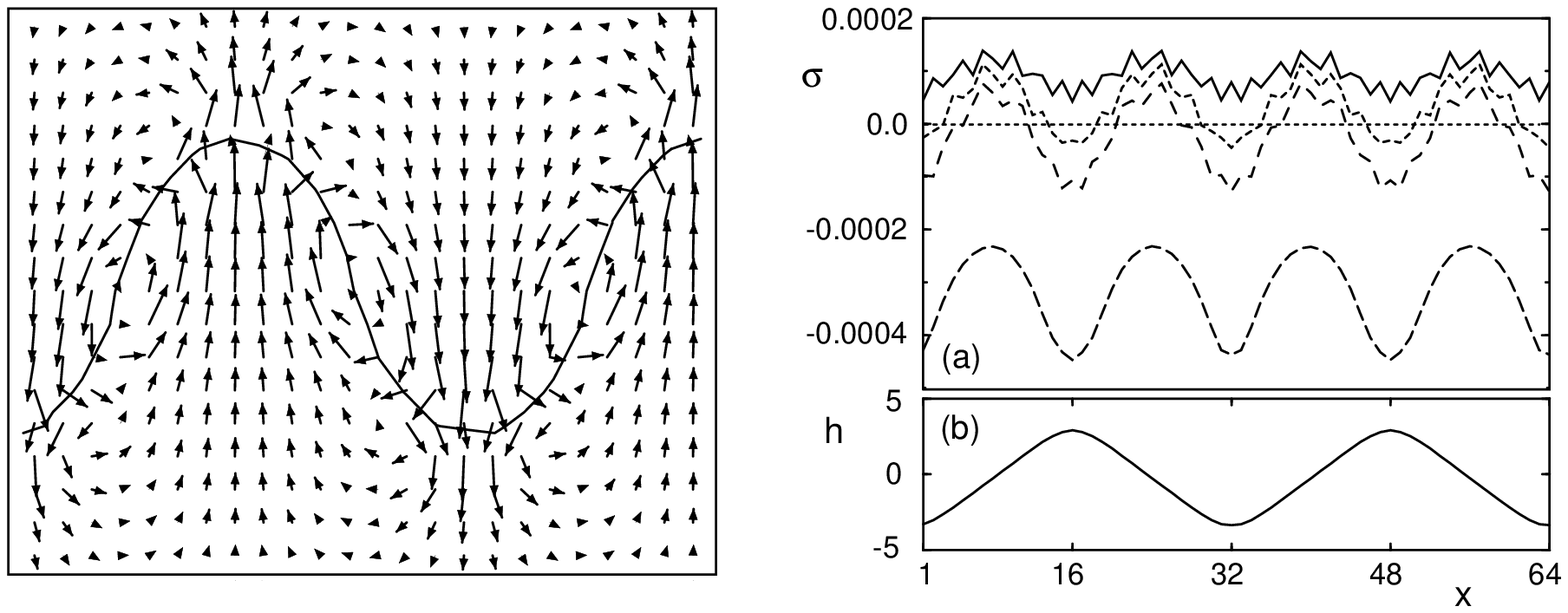}
\caption{
Typical flow pattern some time after the quench, where
a large deformation of the interface has developed.  Only a part of
the full lattice and a subset of the velocity vectors is shown.  The
parameters are the same as in fig.~\ref{fig2}.}
\label{fig3}
\caption{
(a) Variation of the local interfacial tension
$\sigma(x)$ parallel to the interface at various times after the
quench (from top to bottom), as indicated in fig.~\ref{fig2}.  
(b) Interface position $h(x)$ at the last time shown in (a).  The 
parameters are the same as in fig.~\ref{fig2}.}
\label{fig4}
\end{figure}

\section{Spontaneous Emulsification} When a system is taken abruptly
from oil-water coexistence into the region of the phase diagram where
the lamellar phase is stable, the large oil- and water-domains have to
break up and form a intertwined structure on a much smaller length
scale.  To study this process in the LB scheme, we equilibrate a
system of equal oil- and water-content in a {\it metastable} state of
oil-water coexistence --- with a planar interface --- inside the
lamellar regime.  We then disturb the interface position by a
sinusoidal capillary wave of the largest possible wavelength.  The
behavior of the interface is monitored by measuring its width
$w=(1/L)\sqrt{\sum_i h^2(x_i)}$, where $L$ is the number of lattice
points, and $x$ is the coordinate parallel to, and $h(x)$ is the local
distance from, the initial planar interface.  The time dependence of
$w$ is shown in fig.~\ref{fig2}.  The width first decreases due to a
relaxation of the local interface profile, and a decrease in the
amplitude of the capillary wave.  After about $10^5$ time steps, $w$
increases again due to the increase of the amplitude of a mode of half
the wavelength of the original perturbation.  This mode is excited due
to anharmonic terms in an effective interface Hamiltonian which lead
to the coupling of different --- in particular to neighboring ---
undulation modes.  The new mode dominates because its characteristic
frequency is larger than that of the original mode, compare
eq.~(\ref{spectrum}).  This mode grows slowly until, after about
$6\cdot 10^5$ time steps, $w$ becomes comparable to the interface
width.  A typical flow pattern at this time is shown in
fig.~\ref{fig3}.  At later stages of emulsification, the interface
shape becomes very complicated, and a characterization by the width
$w$ is no longer possible.

It is interesting to study the early stages of the emulsification
process in more detail. To do so, we have calculated the {\it local}
interfacial tension, which is given by 
\begin{equation}
\sigma(x) = \int dz \, [P_N({\bf r}) - P_T({\bf r})] \ ,
\end{equation}
where $P_N=P_{zz}$ and $P_T=P_{xx}$ are the normal and tangential
components of the pressure tensor, respectively.  The result is shown
in fig.~\ref{fig4} for different times after the quench.  After the
initial relaxation, the interfacial tension takes a value, which is
{\it positive}, but nearly vanishes.  After $2\cdot 10^5$ time steps,
a periodic structure appears, with a wavelength {\it half} of that of
the undulation mode.  The smallest tensions are found at the maxima of
the function $h(x)$, {\it i.e.}  where the interface motion is the
strongest, while the largest tensions are located at the inflection
points of $h(x)$.  After about $4\cdot 10^5$ time steps, $\sigma(x)$
becomes {\it negative} at its minima.  The growth of the interface
width now accelerates; this is accompanied by a further decrease of
the interfacial tension.

Thus, we find that the surface tension along the
interface becomes {\it inhomogeneous} in the early stages of the
emulsification process.  This indicates that most of the new
interfacial area is created at the points where the interface moves
fastest.

\section{Summary and Conclusions}
We have shown in this paper how the LB scheme for non-ideal fluid
mixtures, which has been introduced in refs.~\cite{orla95,swif96}, can
be extended to describe amphiphilic systems.  We have tested both the
equilibrium and dynamical properties of our model and have found
excellent agreement of the results with complementary theoretical
approaches.  The model has then been used to study the spontaneous
emulsification of oil and water after a quench into the lamellar
phase.

Another conserved density, which we haven't taken into account in
our present model, is the density of the amphiphile itself.
Ginzburg-Landau models, which include this density, have already been
investigated in thermal equilibrium \cite{gg:gomp94d,gg:gomp96d}, so
that the construction of a LB model for this case should be
straightforward.  The diffusion of the amphiphile to the interface
should lead to new dynamical phenomena \cite{gran93}.

\stars

We acknowledge support from the Deutsche Forschungsgemeinschaft, the
National Science Foundation under Grants No.~DMR-9405824 and
DMR-9712134 and the donors of The Petroleum Research Fund,
administered by the ACS.

\bibliographystyle{}

%\bibliography{gompper,amphiphile,preprints}

\end{document}